\documentclass[a4paper,11pt]{article}
\usepackage{pos}

\usepackage{subfigure}

\usepackage{slashed}
\usepackage{mathtools}

\usepackage{feynmp}
\usepackage{feynmp-auto}
\newcommand{\N}{\mathcal{N}}

\title{Brief on Dark Matter in the Type Ib Seesaw Model: a GeV-scale Dirac neutrino portal }

\author*[a]{Bowen Fu}
\emailAdd{B.Fu@soton.ac.uk}

\affiliation[a]{Department of Physics and Astronomy, University of Southampton,\\
SO17 1BJ Southampton, United Kingdom}

\abstract{The type Ib seesaw, as an alternative explanation to the origin of neutrino mass, provides a new intriguing way to connect the neutrino physics to cosmology. In this proceeding, we consider a minimal type Ib seesaw model where the effective neutrino mass operator involves two different Higgs doublets and a heavy Dirac mass. We propose a minimal dark matter extension of this model, in which the Dirac heavy neutrino is coupled to a dark Dirac fermion and a dark complex scalar field, both odd under a discrete $Z_2$ symmetry, where the lighter one serves as a dark matter candidate. Focussing on the fermionic dark matter case, we explore the parameter space of the seesaw Yukawa couplings, the neutrino portal couplings and dark scalar to dark fermion mass ratio, where correct dark matter relic abundance can be produced by the freeze-in mechanism. By considering the mixing between the standard model neutrinos and the heavy neutrino, a connection can be built between dark matter production and laboratory experiments.}

\FullConference{%
  *** The European Physical Society Conference on High Energy Physics (EPS-HEP2021), ***\\
  *** 26-30 July 2021 ***\\
  *** Online conference, jointly organized by Universität Hamburg and the research center DESY ***
}

\begin{document}
\maketitle

\section{Introduction \label{sec:Intro}}

One of the interesting possibilities to relate the dark matter (DM) problem to neutrino mass and mixing is through the so-called neutrino portal where a dark fermion and a dark scatter interact with heavy neutrinos. Recent researches \cite{Chianese:2018dsz,Chianese:2019epo,Chianese:2020khl} show that dark matter particles can be dominantly produced through the neutrino Yukawa interactions in the seesaw sector non-thermally. However, the right-handed neutrinos are required to be at least TeV scale within the framework of the traditional type I seesaw model to make the seesaw Yukawa coupling large enough to play a non-negligible role in dark matter production, otherwise the connection between neutrino physics and dark matter is lost. This motivates studies of DM in the type Ib seesaw model \cite{Chianese:2021toe}, where the neutrino mass is generated by a new type of Weinberg operator involving two Higgs doublets \cite{Hernandez-Garcia:2019uof} and a Dirac heavy neutrino. In such a model, the seesaw couplings can be large enough to drive DM production for GeV scale heavy neutrinos and thus a testable connection between neutrino physics and dark matter problem is provided.

\section{Minimal type Ib seesaw model with dark matter \label{sec:Model}}

\begin{table}[t!]
\centering
\begin{tabular}{|c|c|c|c|c|c|c|c|c|c|c|c|}
\hline 
& ${Q}_\alpha$ & ${u_R}_\beta$ & ${d_R}_\beta$
& ${L}_\alpha$ & ${e_R}_\beta$ & $\Phi_{1}$ & $\Phi_{2}$  & $\N$ & $\phi$ & $\chi$ \\ \hline  
$SU(2)_L$ & {\bf 2} & {\bf 1} & {\bf 1} & {\bf 2} & {\bf 1} & {\bf 2} & {\bf 2} & {\bf 1} & {\bf 1} & {\bf 1} \\ \hline & & & & & & & & & &  \\[-14pt]
$U(1)_Y$ & $\frac{1}{6}$ & $\frac{2}{3}$ & $-\frac{1}{3}$ & $-\frac{1}{2}$ & $-1$ & $-\frac{1}{2}$ & $-\frac{1}{2}$ & 0 & 0 & 0  \\[1.5pt] \hline & & & & & & & & & &  \\[-14pt]
$Z_3$ & $1$ &  $\omega$ &   $\omega$ & $1$ &  $\omega$ &   $\omega$ &   $\omega^2$ & $\omega$ &  $\omega$  & $\omega^2$  \\ \hline  
$Z_2$ & $+$ & $+$ & $+$ & $+$ & $+$ & $+$ & $+$  & $+$ & $-$ & $-$ \\ \hline
\end{tabular}
\caption{\label{tab:matter}Irreducible representations of the fields of the model under the electroweak $SU(2)_L\times U(1)_Y$ gauge symmetry, the discrete $Z_3$ symmetry (where $\omega =e^{i2\pi/3}$) and the unbroken $Z_2$
dark symmetry. The fields $Q_{\alpha}, L_{\alpha}$ are left-handed SM doublets while ${u_R}_\beta,{d_R}_\beta,{e_R}_\beta$
are right-handed SM singlets where $\alpha, \beta = 1,2,3$ label the three families of quarks and leptons. The fields $\N$ is the heavy Dirac neutrino, while $\phi$ and $\chi$ are a dark complex scalar and dark Dirac fermion, respectively.}
\end{table}

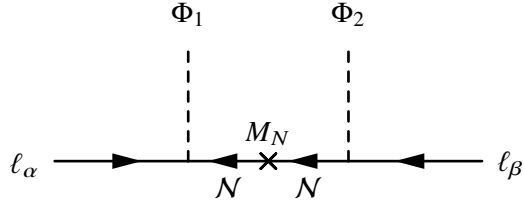
\begin{figure}[t!]
\begin{center}
\begin{fmffile}{FeynDiag/NM}
\fmfframe(10,10)(10,10){
\begin{fmfgraph*}(200,45)
\fmflabel{$\ell_\beta$}{o1}
\fmflabel{}{o2}
\fmflabel{}{i2}
\fmflabel{$\ell_\alpha$}{i1}
\fmfv{}{v1}
\fmfv{decor.shape=cross,decor.size=8,label=$M_N$,label.angle=90}{v2}
\fmfv{}{v3}
\fmfv{label=$\Phi_1$,label.angle=90}{v4}
\fmfv{}{v5}
\fmfv{label=$\Phi_2$,label.angle=90}{v6}
\fmfleft{i1,i2}
\fmfright{o1,o2}
\fmf{fermion,tension=0.6}{i1,v1}
\fmf{fermion,label=$\N$,l.side=left}{v2,v1}
\fmf{fermion,label=$\N$,l.side=left}{v3,v2}
\fmf{fermion,tension=0.6}{o1,v3}
\fmf{phantom,tension=0.6}{i2,v4}
\fmf{phantom}{v5,v4}
\fmf{phantom}{v5,v6}
\fmf{phantom,tension=0.6}{o2,v6}
\fmf{dashes,tension=0}{v1,v4}
\fmf{phantom,tension=0}{v2,v5}
\fmf{dashes,tension=0}{v3,v6}
\end{fmfgraph*}}
\end{fmffile}
\caption{\label{fig:Ib} The type Ib seesaw mechanism involves two different Higgs doublets $\Phi_1$ and $\Phi_2$.}
\end{center}
\end{figure}

Here we introduce the minimal version of the type Ib seesaw model with a heavy Dirac neutrino~\cite{Chianese:2021toe}, 
as shown in Tab.\ref{tab:matter}. All fields transform under a $Z_3$ symmetry in such a way as to ensure that the coupling between the Higgs doublets and SM fermions follows the type II 2HDM pattern and require two different Higgs doublets in the seesaw mechanism. We also consider a dark matter extension of this model to include a $Z_2$-odd dark sector containing a singlet Dirac fermion $\chi$ and a singlet complex scalar $\phi$. The vacuum of the dark scalar field is required to appear at zero in the potential so that the $Z_2$ symmetry is preserved. The type Ib seesaw Lagrangian and the neutrino portal read 
\begin{eqnarray}
\mathcal{L}_{\rm seesawIb} & = & - Y_{1\alpha}^* \overline{L^c}_\alpha \Phi_1^* \N_L - Y_{2\alpha} \overline{L}_\alpha {\Phi}_2 \N_R - M_N\overline{\N_L} \N_R+ {\rm h.c.}\,, 
\label{eq:lagNS}\\
\mathcal{L}^{\rm Parity}_{\rm N_R portal} & = &  y \phi \, \overline{\chi}\N   + {\rm h.c.} \,.
\label{eq:lagPortal}
\label{eq:ds}
\end{eqnarray}
In fact, the vector-like heavy neutrino $\N$ with a Dirac mass $M_N$ can also be understood as a pair of ``right-handed'' Weyl neutrinos $\left( N^c_{R1},N_{R2} \right)$ with an off-diagonal mass matrix~\cite{Chianese:2021toe}.

In the type Ib seesaw model, dimension five Weinberg-type operators involving two different Higgs doublets \cite{Hernandez-Garcia:2019uof} can be obtained as shown in Fig.\ref{fig:Ib}. The new Weinberg-type operator induces Majorana mass terms $m_{\alpha \beta} \nu_{\alpha} \nu_{\beta}$ for the light SM neutrinos after the Higgs doublets develop VEVs
\begin{equation}
m_{\alpha \beta} = \dfrac{  v_1  v_2}{M_N}\left(Y_{1\alpha}^* Y_{2\beta}^*  + Y_{1\beta}^* Y_{2\alpha}^* \right)\,.
\label{eq:dim5_relation_simp}
\end{equation} 
The Yukawa couplings $Y_{1\alpha}, Y_{2\alpha}$ in type Ib seesaw model can be determined the neutrino oscillation data up to overall factors $Y_1$ and $Y_2$~\cite{Hernandez-Garcia:2019uof}. For the case of a normal hierarchy (NH), the Yukawa couplings in the flavour basis, where the charged lepton mass matrix is diagonal, read
\begin{eqnarray}
Y_{1\alpha} &=&\dfrac{Y_1}{\sqrt{2}}\left( \sqrt{1+\rho} \left(U_\text{PMNS}\right)_{\alpha 3}-\sqrt{1-\rho} \left(U_\text{PMNS}\right)_{\alpha 2}\right)\,, 
\label{eq:Y1_NH}\\
Y_{2\alpha} &=& \dfrac{Y_2}{\sqrt{2}}\left( \sqrt{1+\rho} \left(U_\text{PMNS}\right)_{\alpha 3}+\sqrt{1-\rho} \left(U_\text{PMNS}\right)_{\alpha 2}\right)\,, 
\label{eq:Y2_NH}
\end{eqnarray}
where $Y_2,\,Y_1$ are real numbers and  $\rho={(\sqrt{1+r}-\sqrt{r})}/{(\sqrt{1+r}+\sqrt{r})}$ with $r \equiv {\vert\Delta m_{21}^2\vert}/{\vert\Delta m_{32}^2\vert}$. 

\section{Dark matter production \label{sec:DMpro}}
\begin{figure}[t!]
\begin{center}
\subfigure[~Neutrino Yukawa processes]{
\begin{fmffile}{FeynDiag/NS1}
\fmfframe(10,10)(10,15){
\begin{fmfgraph*}(65,30)
\fmflabel{$\phi^*$}{o1}
\fmflabel{$\chi$}{o2}
\fmflabel{$\nu_\alpha,\ell^+_\alpha$}{i2}
\fmflabel{$\phi_1^0,\phi_1^-$}{i1}
\fmfv{label=$y$}{v2}
\fmfv{label=$Y_{1\alpha}$}{v1}
\fmfleft{i1,i2}
\fmfright{o1,o2}
\fmf{plain}{i2,v1}
\fmf{dashes}{v2,o1}
\fmf{fermion}{v2,o2}
\fmf{dashes}{i1,v1}
\fmf{fermion,label=$\N$}{v1,v2}
\fmfdotn{v}{2}
\end{fmfgraph*}}
\end{fmffile}
\begin{fmffile}{FeynDiag/NS2}
\fmfframe(20,10)(10,15){
\begin{fmfgraph*}(65,30)
\fmflabel{$\phi^*$}{o1}
\fmflabel{$\chi$}{o2}
\fmflabel{$\nu_\alpha,\ell^-_\alpha$}{i2}
\fmflabel{${\phi_2^0},\phi_2^+$}{i1}
\fmfv{label=$y$}{v2}
\fmfv{label=$Y_{2\alpha}^*$}{v1}
\fmfleft{i1,i2}
\fmfright{o1,o2}
\fmf{plain}{i2,v1}
\fmf{dashes}{v2,o1}
\fmf{fermion}{v2,o2}
\fmf{dashes}{i1,v1}
\fmf{fermion,label=$\N$}{v1,v2}
\fmfdotn{v}{2}
\end{fmfgraph*}}
\end{fmffile}\label{fig:feynA}}
\subfigure[~Dark sector processes]{
\begin{fmffile}{FeynDiag/DS1}
\fmfframe(10,10)(10,15){
\begin{fmfgraph*}(65,30)
\fmflabel{$\phi$}{o1}
\fmflabel{$\phi^*$}{o2}
\fmflabel{$\overline{\N}$}{i1}
\fmflabel{$\N$}{i2}
\fmfv{label=$y$}{v1}
\fmfv{label=$y$}{v2}
\fmfleft{i1,i2}
\fmfright{o1,o2}
\fmf{fermion}{i2,v1}
\fmf{dashes}{v1,o2}
\fmf{fermion}{v2,i1}
\fmf{dashes}{v2,o1}
\fmf{fermion,label=$\chi$,tension=0}{v1,v2}
\fmfdotn{v}{2}
\end{fmfgraph*}}
\end{fmffile}
\begin{fmffile}{FeynDiag/DS2}
\fmfframe(10,10)(10,15){
\begin{fmfgraph*}(65,30)
\fmflabel{$\overline{\chi}$}{o1}
\fmflabel{$\chi$}{o2}
\fmflabel{$\overline{\N}$}{i1}
\fmflabel{$\N$}{i2}
\fmfv{label=$y$}{v1}
\fmfv{label=$y$}{v2}
\fmfleft{i1,i2}
\fmfright{o1,o2}
\fmf{fermion}{i2,v1}
\fmf{fermion}{v1,o2}
\fmf{fermion}{v2,i1}
\fmf{fermion}{o1,v2}
\fmf{dashes,label=$\phi$,tension=0}{v1,v2}
\fmfdotn{v}{2}
\end{fmfgraph*}}
\end{fmffile}
\label{fig:feynB}}
\end{center}
\caption{\label{fig:Feyn} 
Processes responsible for the dark matter production considered in this study. 
}
\end{figure}
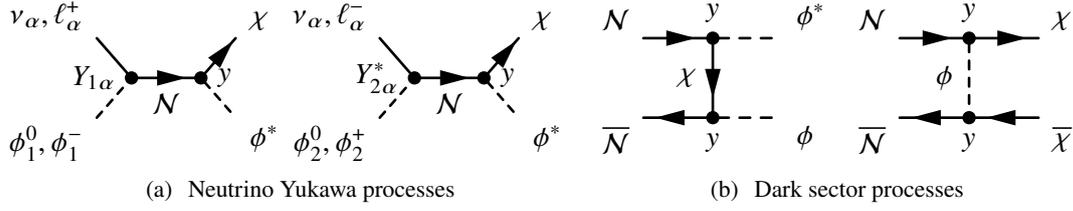

We focus on the freeze-in production of dark matter in which scenario the dark scalar can decay into the dark fermion and heavy neutrinos. The Feynman diagrams for processes that are relevant to dark matter production are shown in Fig.\ref{fig:Feyn}, which are classified into the neutrino Yukawa processes and the dark sector processes. 
By solving the Boltzmann equations, the constraint from observation result by the Planck Collaboration at 68\% C.L.~\cite{Aghanim:2018eyx} on the couplings can be obtained as
\begin{eqnarray}
y^2\left( Y_1^2 + Y_2^2 \right) \simeq 2.1 \times 10^{-22} {m_\phi}/{m_\chi}\, \quad\text{and}\quad
y^4 \simeq 3.9 \times 10^{-23} {m_\phi}/{m_\chi}\, \label{eq:couplings}
\end{eqnarray}
when the $\nu$-Yukawa processes and the dark sector processes dominate the dark matter production, respectively. For heavy Dirac neutrinos at 1-100 GeV scale, by applying the relation in Eq.\eqref{eq:couplings} to the active-sterile neutrino mixing strength, we found that the parameters in the model can be constrained by the current experimental results and tested by future experiments \cite{SHiP:2018xqw,Blondel:2014bra}, especially when the ratio of the Higgs VEVs is small and when the ratio of dark particle masses is large \cite{Chianese:2021toe}. 

\section{Conclusion \label{sec:Concl}}

We have studied the dark matter production in the simple dark matter extension of the minimal type Ib seesaw model, with the heavy Dirac neutrino portal to a dark scalar and a dark fermion. Due to the special structure of the type Ib seesaw model, the parameters in the model are highly constrained by the oscillation data and the dark matter production has an interesting dependence on the seesaw Yukawa couplings. By considering the neutrino mixing, the relation of parameters obtained from DM production can be related to collider experiments. However, this is not the only connection between cosmological problems and the type Ib seesaw mechanism as it has been pointed out that the model can produce baryon asymmetry in a variant version \cite{Fu:2021fyk} as well as include another DM candidate through a gauge extension \cite{Fu:2021uoo}. The type Ib seesaw model not only provides an alternative explanation to the origin of neutrino mass but also opens another gate from neutrino physics to cosmological mysteries.

\bibliographystyle{JHEP}
\bibliography{TypeIb}

\end{document}